\documentclass[aps,pra,twocolumn,groupedaddress,showpacs,amsmath,amssymb]{revtex4}
\usepackage{graphicx,xspace}

\begin{document}

\newcommand{\rcm}{\mbox{cm$^{-1}$}\xspace}
\newcommand{\cms}{\mbox{cm$^{3}$s$^{-1}$}\xspace}
\newcommand{\OH}{OH$^-$\xspace}
\newcommand{\NE}{Ne$^+$\xspace}
\newcommand{\RB}{Rb$^+$\xspace}

\title{Reactive collisions of trapped anions with ultracold atoms}

\author{J. Deiglmayr}
\altaffiliation{Permanent address: Laboratory of physical chemistry, ETH Z\"urich}
\author{A. G\"oritz}
\altaffiliation{Permanent address: Physikalisches Institut, Albert-Ludwigs-Universit\"at Freiburg}
\author{T. Best}
\author{M. Weidem\"uller}
\altaffiliation{Permanent address: Physikalisches Institut, Ruprecht-Karls-Universit\"at Heidelberg}
\author{R. Wester}
\email{roland.wester@uibk.ac.at}

\affiliation{Institut f\"ur Ionenphysik und Angewandte Physik, Universit\"at Innsbruck, Technikerstra\ss{}e 25/3, 6020 Innsbruck, Austria}

\date{\today}

\begin{abstract}
We present a scheme to embed molecular anions in a gas of
ultracold rubidium atoms as a route towards the preparation of cold
molecular ions by collisional cooling with ultracold atoms. Associative detachment as an important loss process in collisions between OH$^-$ molecules and rubidium atoms is studied. The density distribution of trapped negative ions in the multipole radiofrequency trap is measured by photodetachment tomography, which allows us to derive absolute rate coefficients for
the process. We define a regime where translational and internal cooling of molecular ions embedded into the ultracold atomic cloud can be achieved.
\end{abstract}

\maketitle

Cold molecules have stimulated a lot of research in the last decade
since they offer new opportunities for ultrahigh precision spectroscopy, for studying ultracold gases with anisotropic interactions, and for ultracold chemistry driven by quantum mechanics~\cite{Carr2009,Dulieu2009,schnell2009}. This has created a new research field that has focussed on the formation, preparation and control of cold molecules with translational, rotational, vibrational and hyperfine energies corresponding to temperatures below 1~Kelvin. Most of these studies are based on specialized preparation techniques to achieve molecular temperatures far below liquid helium temperature, based on the molecular Stark shift, magnetic moment, or the association of laser-cooled atoms.

For molecular ions the situation is advantageous, because collisional cooling is a very general method to cool ions~\cite{Schiller2009}. Translational temperatures down to few millikelvin are provided by sympathetic cooling of molecular ions with laser-cooled atomic ions trapped jointly in a radiofrequency (rf) ion trap. However, ion-ion collisions do not lead to internal quantum state cooling due to the long-range nature of the interaction. Therefore complex and molecule-specific optical pumping or state-selected photoionisation schemes have to be used for rotational state control~\cite{Schneider2010, Staanum2010,Tong2010}. Alternatively, molecular ions can be cooled in collisions with cold neutral atoms, which feature much closer encounters and thereby allow for efficient cooling of rotational and vibrational states in inelastic collisions. This general cooling of all molecular degrees of freedom can be applied to molecular ions and ionic clusters of almost arbitrary size and complexity. However, it limits experiments to temperatures above about 4\,K due to the limitations imposed by standard cryostats.

In this letter we present a hybrid ion-atom trap for the study of interactions of trapped molecular ions with ultracold atoms. We discuss the prospects of this trap for collisional cooling of complex molecular ions to temperatures below 1 Kelvin. In contrast to several recent experiments that study the interaction of single atomic cations with ultracold atoms~\cite{Smith2005,Grier2009,Zipkes2010,Schmid2010,Hall2011},
we do not employ a Paul trap, but an octupole ion trap. This strongly suppresses disturbances of ion-atom collisions due to micro-motion throughout the extended volume of the atom trap~\cite{Wester2009,Asvany2009}. As a spectroscopic diagnostics of the ions' density distribution, which can be related to their translational temperature, we employ single-photon photodetachment spectroscopy~\cite{Trippel2006}. This is a general method that can be applied to all negative ions and, while species- and state-selective, does not depend on the existence of a closed cycling transition as it is the case for fluorescence or absorption imaging.

Certain conditions have to be fulfilled for efficient collisional cooling of an ion by a neutral. First, as in collisions between neutrals, the system has to exhibit a large ratio of elastic or rotationally inelastic collisions to reactive collisions, which lead to a loss of the original ion. A prominent channel of such reactive collisions, which hinder cooling, is charge transfer from the ion to the neutral coolant, as observed in recent experiments involving cationic atoms~\cite{Schmid2010,Hall2011}. A recent detailed investigation of collision processes in the \OH + Rb system revealed very promising conditions for collisional cooling of internal and external degrees of freedom of the molecular anion by ultracold rubidium~\cite{Gonzalez2008,Tacconi2009}. Furthermore charge transfer from \OH to rubidium is energetically forbidden as the electron affinity of OH greatly exceeds the one of Rb (1.83eV~\cite{Smith1997} versus  0.486~\cite{Frey1978} respectively). There remains an important loss mechanism, namely associative detachment (AD)
\begin{equation}
\textrm{OH}^- + \textrm{Rb} \rightarrow \textrm{RbOH} + \textrm{e}^- \label{eq:defAD}
\end{equation}
which is exothermal by 1.4~eV~\cite{Smith1997,Lara2006}. To our knowledge the cross section for this process has not been investigated yet, however early studies of associative detachment involving \OH showed significant rate constants close to the capture limit~\cite{Howard1974}. Here we measure the rate coefficients for the reactive losses of the molecular ion \OH due to collisions with ultracold rubidium and use this result to define a regime where translational and internal cooling of molecular ions embedded into the ultracold atomic cloud can be achieved.

\begin{figure}[bt]
\includegraphics[width=0.9 \columnwidth]{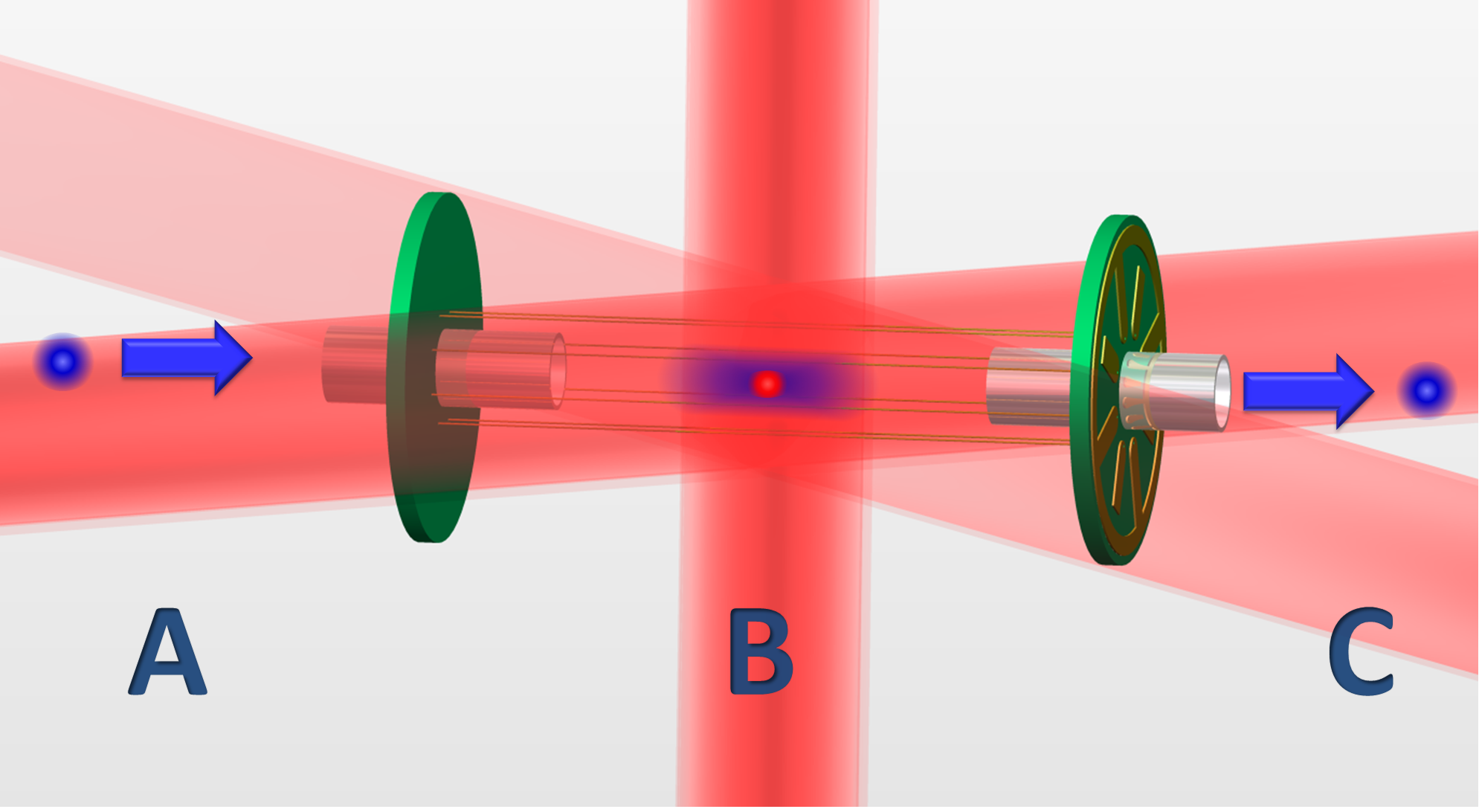}
\caption{Schematic depiction of the hybrid atom-ion trap. (A) anions are produced by a plasma-discharge in a gas jet emerging from a pulsed valve, mass-selected by time-of-flight, and loaded into the trap through a hollow end cap. (B) The ion trap consists of eight parallel 100~$\mu$m thin wires forming an octupole to which sinusoidal voltages with a frequency of 7.7~MHz and amplitudes around 150~V are applied. Longitudinal confinement is provided by two end caps separated by 34~mm to which voltages around 20~V are applied. A cloud of 10$^7$ rubidium atoms is laser-cooled and trapped in the center of the ion cloud. (C) Ions are extracted through the hollow end cap and detected in a time-of-flight mass-spectrometer.}
\label{fig:HAIsetup}
\end{figure}

The setup combines an octupole rf trap for ions with a magneto-optical trap (MOT) for rubidium atoms. A schematic sketch of the trap is shown in figure \ref{fig:HAIsetup}. The rf trap consists of eight parallel wires arranged equidistantly on a circle with a diameter of 6\,mm to which rf is applied. This rf field creates an effective potential along the radial dimension $r$ which is almost flat in the center and exhibits steep walls proportional to $r^6$~\cite{Wester2009}. Confinement along the trap axis is provided by two hollow end caps.  A simulation of the ion density at room temperature in this trapping potential with typical parameters for rf amplitude (150~V) and end cap voltage (20~V) predicts in axial direction a Gaussian distribution with a full-width at half maximum (FWHM) of 4.0~mm and in radial direction an almost uniform distribution with a FWHM of 2.0~mm. The volume occupied by laser-cooled rubidium atoms (typical FWHM 1.0~mm) is thus fully embedded in the ion cloud. Furthermore the largest value of the effective potential in which collisions between ions and neutrals can occur corresponds to few Kelvin. This significantly limits the maximum amount of energy transferred from the electric field to the ions kinetic energy during a collision event and thus limits rf heating \cite{Wester2009,Asvany2009}. It also suppresses losses due to collisions under non-adiabatic conditions in high rf amplitudes~\cite{Mikosch2007}.

Ions are created from neutral precursors in a supersonic expansion through a plasma discharge, mass selected, accelerated towards the trapping region, and eventually loaded into the rf trap along the trap axis through the hollow entrance cap. During loading neon gas is fed into the trapping chamber through a pulsed valve which facilitates initial loading and thermalization of the ion cloud.
During storage a background pressure of typically $5\cdot10^{-9}$~mbar is maintained. For detection the ions are accelerated through the exit endcap towards a channeltron detector. A time-of-flight tube of 500\,mm length between trap and detector allows for distinguishing ions with different ratios of $q/m$.

In order to combine the ion trap with laser cooled atoms, the rf trap is mounted between two coils (free distance 50mm) which create the magnetic field gradient for the MOT. The light for laser cooling of rubidium atoms (collimated waist 1\,cm, total peak light intensity 38mW/cm$^2$) passes in the z-direction (perpendicular to the ion trap axis) through the center of the magnetic coils. In the orthogonal plane the beams cross the axis of the rf trap under an angle of 30$\,^{\circ}$ each. Rubidium atoms are provided by two rubidium dispensers mounted at a distance of roughly 5\,cm from the trap center where the line-of-sight from the hot dispensers to the trap center is blocked by a wire. The size and position of the MOT is monitored by recording the fluorescence of the trapped atoms via two CCD-cameras under an angle of 90$\,^{\circ}$. The magnification of the imaging system is obtained by observing the light scattered from the wires of the rf trap with well known spacings. The number of trapped rubidium samples is then calculated from the measured dimensions of the atom cloud assuming a typical atom  peak density of $2\cdot10^{10}$ atoms/cm$^3$.

The spatial distribution of ions in the trap is determined via photodetachment tomography~\cite{Trippel2006,Hlavenka2009}. In contrast to established imaging techniques for gases of neutral particles, such as absorption imaging or fluorescence detection, photodetachment is a single-photon effect which does not rely on closed cycling transitions and can thus be applied to any trapped anion. Also photodetachment close to threshold can be used to probe the internal state distribution of  trapped anions~\cite{Zastrow2012}. For photodetachment of \OH we use light from a free running diode laser at 660~nm, well above the threshold at 678~nm, focused with a cylindrical lens to create a line focus with axial width of 92\,$\mu$m and almost homogeneous intensity distribution over the ion trap in vertical direction. This line focus is then scanned along the trap axis. The photodetachment process is observed via the light-induced trap loss, which is proportional to the column density of ions along the beam propagation.

\begin{figure}[tb]
\includegraphics[width=0.9 \columnwidth]{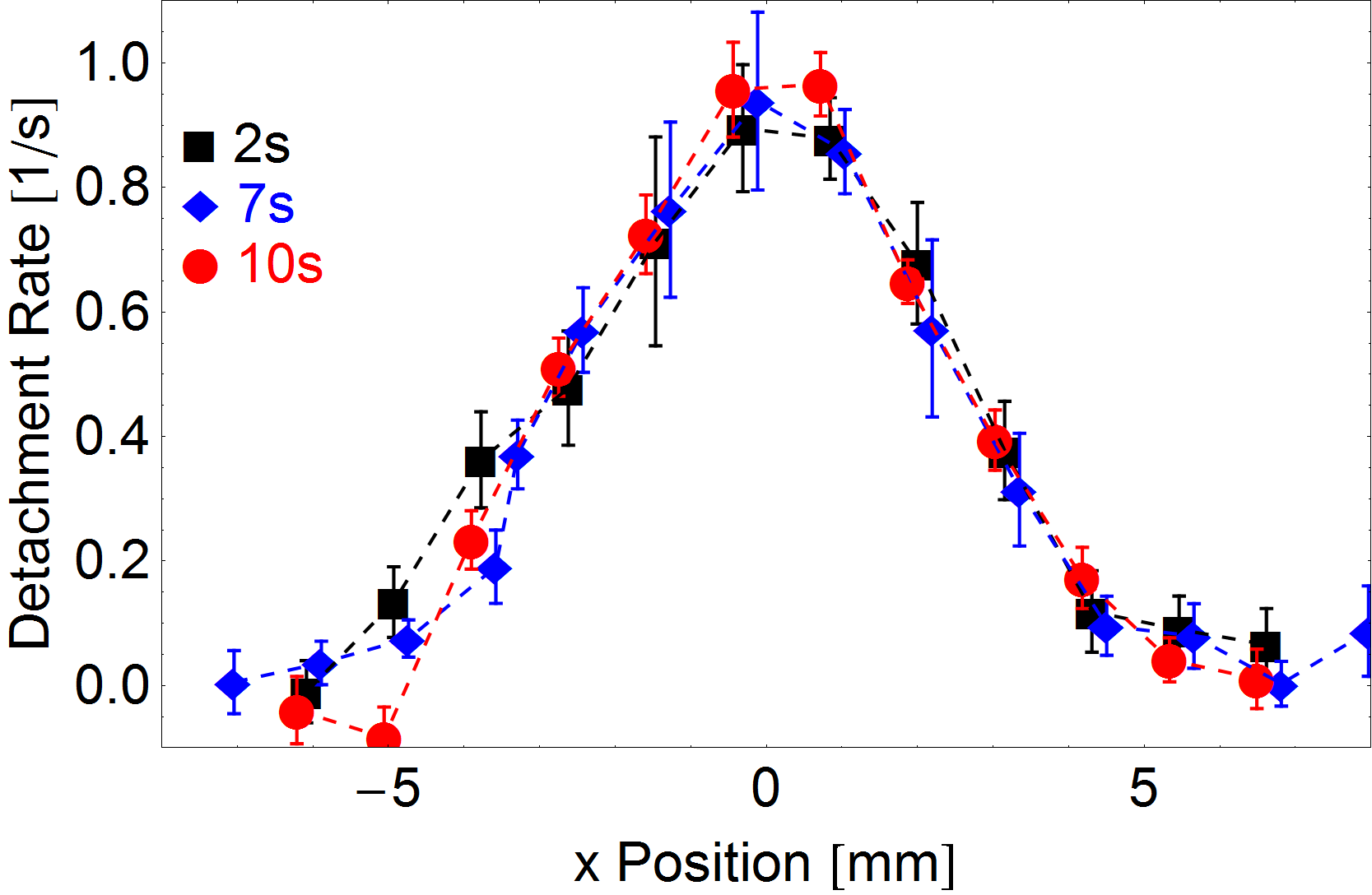}
\caption{(Color online) Density profiles of \OH ions in the trap after three interaction times with rubidium. Shown are the photodetachment rates at different trap positions, which are proportional to the local column density along the laser beam. Each point is determined from fitting a first-order decay process to the trapped ion signal during 2\,s interaction time with the detachment laser. Given errorbars represent the standard deviation of the fit procedure. Note that the MOT is only fully loaded after roughly 2~s.}
\label{Fig:Thermometry}
\end{figure}

Measurements of the ion distribution along the trap axis are shown in fig.~\ref{Fig:Thermometry} for \OH ions stored together with Rb atoms for storage times between 2 and 10s. The widths of these distributions measure the translational temperature of the ions. Specifically, as the axial confinement is well described by a harmonic potential, the ion temperature scales with the square of the axial width. The absolute kinetic temperature is estimated from a simulation of the effective trapping potential and the resulting ion density distribution to be $470\pm200$K, where the accuracy is mostly limited by counting statistics and by the precision of the electric field simulation. It can be seen that irrespective of the interaction time with the Rb atoms, no change in translational temperature is observed. We estimate our method to be sensitive to relative changes exceeding 20\%. With the given Rb atom number, \OH ion distribution, and the Langevin rate-coefficient for \OH-Rb collisions we estimate a collision rate on the order of 8\,s$^{-1}$, which should suffice for a significant cooling of the trapped ions within about a second. The absence of cooling is attributed to competing thermalization, most likely due to collisions with the residual background gas. While the latter is about two orders of magnitude lower in density, it overlaps the ions throughout the trap whereas the Rb atoms fill only a small volume in the trap center. In particular resonant proton transfer with H$_2$O molecules, known to be a significant fraction of the residual gas due to its use in the ion source, would lead to room temperature \OH in just a single collision. The rate for this process is estimated to be about 0.4\,s$^{-1}$~\cite{Maergoiz2009}. Following from simple kinematic arguments, elastic \OH-Rb collisions should lead to an exponential decay of the ion temperature with a time constant of 3\,s, thus on the same time scale as the heating due to collisions with background gas.

\begin{figure}[tb]
\includegraphics[width=0.9 \columnwidth]{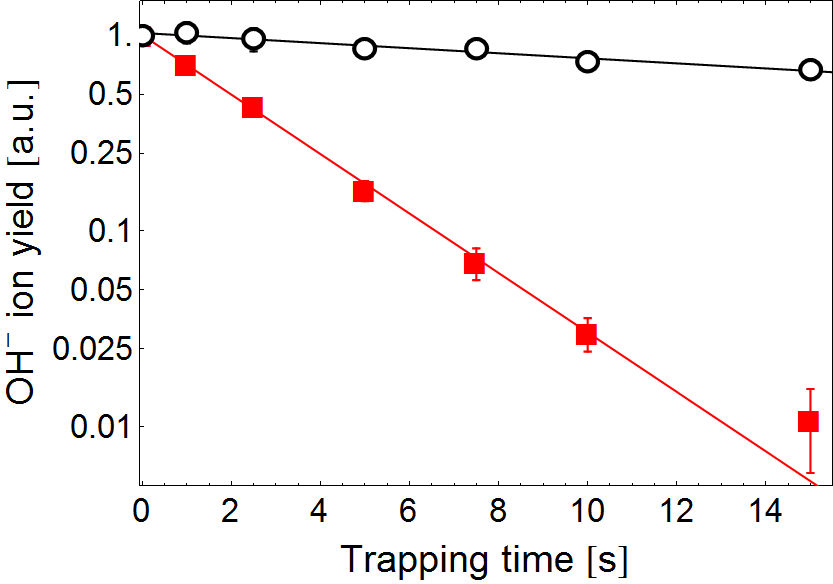}
\caption{(Color online) Detected \OH ions after simultaneous trapping of ions and laser cooling of rubidium atoms (red squares). Also shown is the reference measurement (black circles) without trapped rubidium atoms (see text for details).}
\label{Fig:OHRbLoss}
\end{figure}

\begin{figure}[tb]
\includegraphics[width=0.9 \columnwidth]{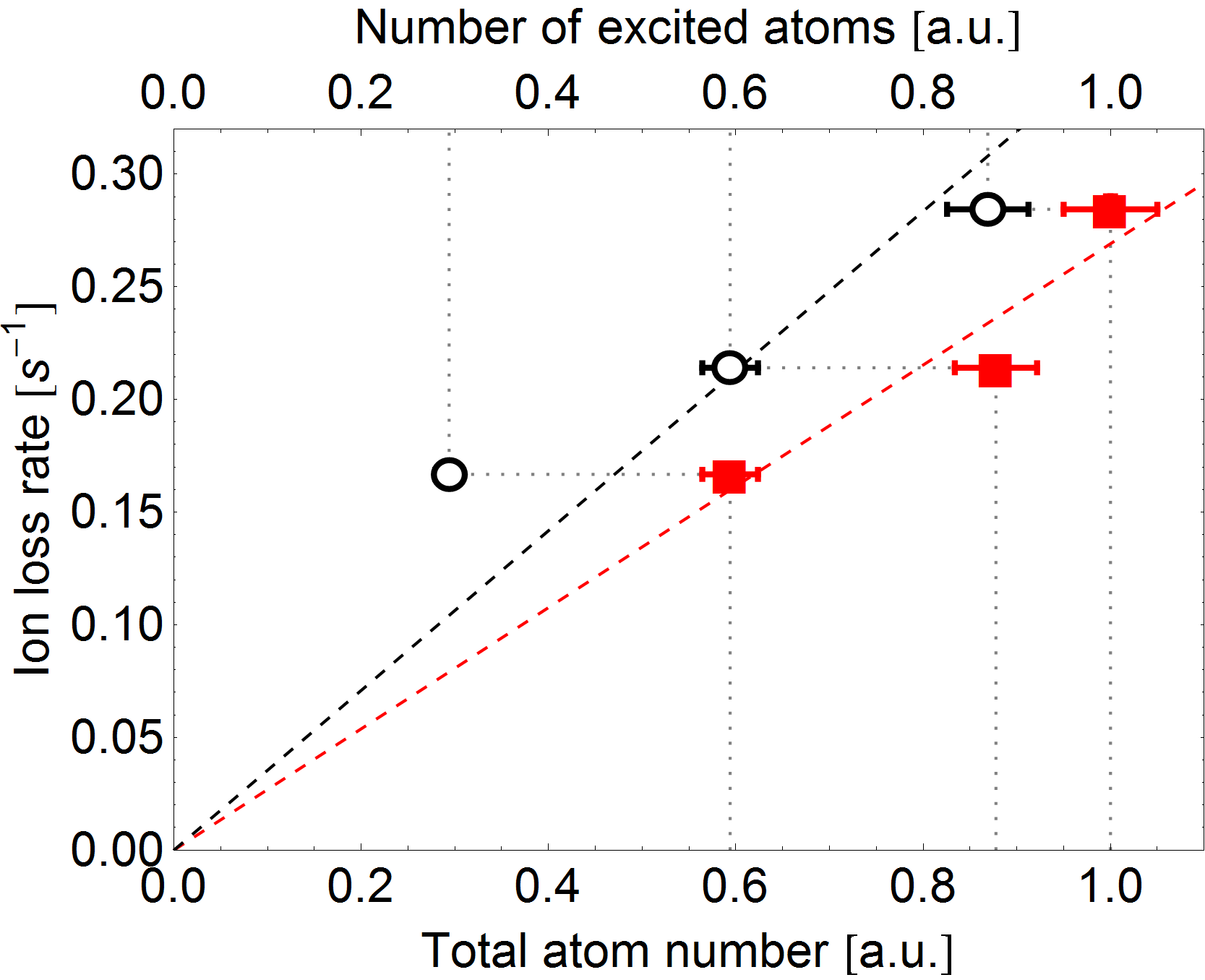}
\caption{(Color online) Measured loss rate for different MOT detunings (3.4$\Gamma$, 2.7$\Gamma$, and 2.0$\Gamma$). The three loss rates are plotted once against the total number of atoms (red squares, determined from the amount of fluorescence light at a fixed detuning) and once against the number of atoms in the excited state (black circles, determined from the amount of fluorescence light at the given MOT detuning). The dotted lines connect each data set, the dashed lines are fits to a linear model.}
\label{Fig:LossMOT}
\end{figure}

Strong losses of \OH ions from the trap are induced by the presence of ultracold rubidium atoms. Figure~\ref{Fig:OHRbLoss} shows the number of trapped \OH ions during trapping. When the cooling laser is tuned to the blue of the cycling transition, and thus no rubidium atoms are trapped, we observe a loss of \OH ions with a time constant of $47\pm3$\,s. When the cooling laser is detuned to the red of the cooling transition by 2.2$\Gamma$, about $4\cdot10^{7}$ atoms are trapped in the center of the ion cloud and induce an ion loss with a rate of $0.329\pm0.018$ 1/s. Due to the large depth of the ion trap and the low temperature of the rubidium atoms the kinetic energy transfer in elastic collisions can not lead to significant ion losses.  Thus the observed losses can be attributed dominantly to inelastic collisions between \OH ions and ultracold $^{85}$Rb atoms. As pointed out in the introduction, the process of associative detachment (eq.~\ref{eq:defAD}) is energetically allowed. Another possible reaction channel is $\textrm{OH}^-+\textrm{Rb}^*\rightarrow \textrm{OH}^-+\textrm{Rb}+E_\textrm{kin}$, a collision between a rubidium atom in the excited state and \OH, potentially releasing 1.6\,eV of energy which would suffice to eject the ion out of the trap. At a given time roughly half of the atoms in the MOT are in the excited state, thus both processes might contribute. In order to clarify the influence of excited rubidium atoms in the MOT we vary the cooling laser detuning and thus the excited state fraction from 25\% to 44\%. The observed loss rates shown in figure~\ref{Fig:LossMOT} exhibit a stronger correlation with the absolute number of atoms than with the number of excited atoms, which suggests a dominant contribution from AD in the ground state.


The rate constant for the observed inelastic losses is derived from the independent determination of ion and atom distributions. As shown above we can assume that the rubidium sample is fully embedded in the ion cloud and that the ion density is roughly constant over the MOT region (a detailed numerical simulation confirmed this assumption to be valid with 10\% of the final rate constant). We took care to perform the loss measurements under conditions where a correlation between number of ions and ion density due to Coulomb repulsion can be excluded. The rate constant for inelastic losses $K_\textrm{inel}$ is then given by $K_\textrm{inel} = \frac{R_\textrm{inel}}{ N_{\textrm{Rb}} \rho_\textrm{{0,ion}}}$ where $R_\textrm{inel}$ is the measured rate of inelastic losses, $N_{\textrm{Rb}}$ is the total number of rubidium atoms and $\rho_\textrm{{0,ion}}$ is the peak value of the normalized ion distribution. $N_{\textrm{Rb}}$ is determined as described above from the analysis of fluorescence images, $\rho_\textrm{{0,ion}}$ is calculated from a fit of the trapping potential, which was derived from a numerical simulation, to the experimentally measured widths in axial and radial directions. Performing this analysis for the measurement shown in figure~\ref{Fig:OHRbLoss} we find an inelastic rate coefficient of $(2^{+2}_{-1})\cdot10^{-10}$\,\cms where the uncertainty is mostly caused by a systematic uncertainty in the determination of the atom number. This rate coefficient holds for a relative temperature of the \OH-Rb system of $400\pm200$K, based on the estimated kinetic temperature of \OH and the negligible kinetic temperature of the Rb atoms. This loss rate is significantly smaller than the rate coefficient of $4.3\cdot10^{-9}$\,\cms given by Langevin theory for collisions between \OH and Rb 5S$_{1/2}$ atoms. As AD leads to losses but not to a heating of the remaining ensemble, cooling through elastic collisions should still be possible if the interaction time between ion and neutrals can be limited, \textit{e. g.} by shuttering the MOT after a certain interaction time.

From our observations we derive the following protocol for collisional cooling of anions by ultracold atoms in a rf multipole trap. First, as cooling requires a sufficient number of elastic collisions before inelastic losses can occur, negative ions with a sufficiently small associative detachment rate, suitably an order of magnitude smaller than for \OH+Rb, are required. AD will occur for all anions for which an ion pair state with the cation of the neutral coolant exists and the ion pair can approach close enough to overcome the ionization potential of the coolant atom and couple to the electron continuum. For example, hydrated \OH(H$_2$O)$_\textrm{n}$ clusters have been found to generally show significantly reduced AD rates~\cite{Howard1974} which is explained by the stronger screening of the negative charge by the surrounding water molecules. Also for more delocalized electron wave functions in other polyatomic molecular ions, such as benzene anions or SF$_6^-$, we expect lower AD rates than in the present study. Second, residual collision rates have to be limited far below 0.4 s$^{-1}$ and the rf amplitude has to be reduced following the cooling process to further suppress heating. Third, the interaction between the ions and the atoms has to be limited to a time comparable to the time scale for AD to avoid excessive losses. Furthermore, electronically excited atoms should be removed from the trap using a dark spontaneous force optical trap~\cite{Ketterle1993} to exclude loss processes due to these excited atoms. Following these steps, collisional cooling of a large class of molecular ions to temperatures far below liquid helium temperature seems to become feasible.

\end{document}